\documentclass[12pt,preprint]{aastex}

\begin{document}

\title{Chondrule-Forming Shock Fronts in the Solar Nebula:
A Possible Unified Scenario for Planet and Chondrite Formation}

\author{A. P.~Boss$^1$ and R. H. Durisen$^2$}
\affil{$^1$Department of Terrestrial Magnetism, Carnegie Institution of
Washington, 5241 Broad Branch Road, NW, Washington, DC 20015-1305}
\affil{$^2$Department of Astronomy, Indiana University, 727  E. 3rd St.,
Bloomington, IN 47405-7105}

\authoremail{E-mail: boss@dtm.ciw.edu (A.P.B.); durisen@astro.indiana.edu
(R.H.D.)}

\vspace{1.0in}

\begin{abstract}

Chondrules are mm-sized spherules found throughout primitive, chondritic
meteorites. Flash heating by a shock front is the leading explanation of
their formation. However, identifying a mechanism for creating shock fronts
inside the solar nebula has been difficult. In a gaseous disk capable of
forming Jupiter, the disk must have been marginally gravitationally
unstable at and beyond Jupiter's orbit. We show that this instability can
drive inward spiral shock fronts with shock speeds of up to $\sim$ 10 km
s$^{-1}$ at asteroidal orbits, sufficient to account for chondrule formation.
Mixing and transport of solids in such a disk, combined with the 
planet-forming tendencies of gravitational instabilities, results in a unified
scenario linking chondrite production with gas giant planet
formation.

\end{abstract}

\keywords{solar system: formation -- planetary systems -- accretion,
accretion disks}

\section{Introduction}

 Perhaps the most long-standing problem in all of meteoritics
(Sorby 1877; Hewins 1996) is
the search for an appropriate heat source for melting the chondrules that
constitute the bulk of many primitive meteorites. Chondrules, cm-sized
refractory inclusions, and micron- and smaller-scale matrix particles
came together to form the chondritic meteorites. These three components
appear to have been the major solid constituents of the solar nebula,
at least inside Jupiter's orbit. Their formation and
assembly into chondrites provides strong constraints on
processes occurring during the earliest phases of Solar System origin.

 The chondrule precursors are thought to have been aggregates of
ferromagnesian silicate dust grains that were heated to their liquidus
temperatures through multiple flash heating events, as witnessed by
their textures, inferred cooling rates, occurrence of compound
chondrules, and the occasional presence of fragments of relict
chondrules inside other chondrules (Hewins 1996). Shock waves within the
solar nebula are one possible means for accomplishing this thermal processing
(Hood \& Horanyi 1991, 1993). Detailed studies have shown that such waves are 
able to melt suitable chondrule precursors provided that the shock waves move 
with respect to the precursor aggregates at speeds in the range of 6 to 9 km 
s$^{-1}$ (Iida et al. 2001; Desch \& Connolly 2002; Ciesla \& Hood 2002).

 In this paper we present the results of numerical hydrodynamical models
that identify a likely source for shock heating of chondrule precursors.
We then use these results to create
a unified scenario for the thermal processing of chondrules and mixing
of chondrules, refractory inclusions, and matrix grains to form the
chondritic meteorites.

\section{Sources of Shock Fronts}

A suitable mechanism for producing shock fronts inside the solar nebula
must be identified for the chondrule formation question to be
considered answered. Four different mechanisms have been proposed
as sources of such shock fronts. First, pre-existing dust grains would
be slowed down and heated by frictional gas drag as they enter the
solar nebula by passing through the accretion shock at the nebula's
surface (Wood 1984; Ruzmaikina 1994).
However, this process does not explain the
presence of relict grains in chondrules, or the unlikely growth to
mm-size (or larger, for fluffy aggregates) of grains in the relatively
low density, earlier phases of presolar and interstellar cloud evolution.
Second, the accretion of gas by the solar nebula may have been
punctuated by the infall of clumps of gas and dust whose impact
would have launched shock waves into the disk (Boss \& Graham 1993;
Hood \& Kring 1996). Models
have found that clumps need to have masses of $\sim 10^{27}$ g or
more to produce the desired strong shock waves (Tanaka et al. 1998).
Infalling
clumps with masses greater than this have been inferred in the B335
molecular cloud (Velusamy, Kuiper, \& Langer 1995),
but these clumps may be too distended to
produce an impulsive, shock-forming impact. Third, bow shocks driven
by 1000-km-size planetesimals moving on highly eccentric orbits are another
possibility (Hood 1998; Ciesla, Hood, \& Weidenschilling 2004).
However, achieving such highly eccentric
orbits in the asteroid region requires resonant gravitational interactions
with a Jupiter-like planet (Weidenschilling, Marzari, \& Hood 1998).
In the standard picture
of giant planet formation by core accretion, Jupiter takes several
million years (Ma) or more to form (Inaba, Wetherill, \& Ikoma 2003),
whereas some chondrules
were formed at essentially the same time as the earliest solids
(the calcium, aluminum-rich inclusions, or CAIs) were formed (Bizzarro,
Baker, \& Haack 2004).

 The fourth mechanism is shock fronts associated with nonaxisymmetric
disk structures such as spiral arms (Hood \& Horanyi 1991; Wood 1996). This 
possibility has been strongly supported by recent calculations of the evolution
of gravitationally unstable disks (Boss 2002; Pickett et al. 2003).
The solar nebula is likely to have been at least marginally gravitationally 
unstable during a portion of its evolution, because, regardless 
of mechanism, the formation of Jupiter requires
the mass of the solar nebula to be $\sim 0.1 M_\odot$. Recent models of
Jupiter formation by the core accretion mechanism
require a disk with a mass of at least 0.08 $M_\odot$ in order to
form Jupiter within 3.8 Ma (Inaba et al. 2003). With outer disk
temperatures of $\sim$ 28 K (Kawakita et al. 2001), such a disk would be 
marginally gravitationally
unstable to the growth of density perturbations that would form
strong spiral arms and possibly lead to the rapid formation of Jupiter
(Boss 1997, 2002). The question then becomes whether a disk which
is gravitationally unstable at and beyond Jupiter's 5.2 AU orbital
distance is able to drive
shock fronts into the asteroidal regions ($\sim 2.5$ AU) where
they could have been responsible for {\it in situ}
thermal processing of solids.

\section{Numerical Methods}

 We present here the results of a three dimensional, gravitational
hydrodynamics calculation of the time evolution of a gravitationally
unstable disk which directly addresses this question. 
The calculation includes a full treatment of disk thermodynamics
and radiative transfer in the diffusion approximation. The numerical code
solves the equations of hydrodynamics on a spherical coordinate ($r, \theta,
\phi$) grid with 112 radial grid points spread between 2 AU and 20 AU,
23 $\theta$ grid points between the midplane and the rotation axis,
concentrated around the midplane to provide adequate vertical spatial
resolution, and 256 azimuthal grid points uniformly spaced in $\phi$.
The calculation is identical to previous calculations (e.g., Boss 2002),
except for being extended inward from 4 AU to 2 AU.
The Poisson equation for the gravitational
potential is solved by a spherical harmonic ($Y_{lm}$)
expansion including terms up to $l,m = 32$. The disk has a mass of
$0.096 M_\odot$ between 2 AU and 20 AU. The initial outer
disk temperature is 40 K, leading to an initial value of the Toomre
$Q$ stability parameter of $Q_i \sim 1.3$ beyond 5 AU. Studies of
gravitational instabilities in disks generally show that disks are
unstable to development of nonaxisymmetric structure for
$Q_i <$ 1.5 to 1.7 (e.g., Boss 2002, Pickett et al. 2003). The inner
regions have $Q_i > 1.7$, indicating stability to density perturbations,
because the initial disk has a radial temperature profile that
rises to 1200 K at 2 AU. Hence any spiral structures that might form
in the inner disk must be driven by gravitational forces
from nonaxisymmetric structures in the outer disk.

\section{Results}

 A common outcome of the evolution of gravitationally unstable disks is
the formation of strong, transient shock fronts in the inner disk,
driven by spiral arms and clumps at greater distances (Fig.~1).
For one-armed ($m = 1$ modes) spiral arms of the type seen in
Fig.~1, the inner Lindblad
resonance occurs at angular velocity $\Omega = \infty$, or zero radius, so
that a single clump can drive spiral waves right down to
the protostar. The disk model exhibits this behavior,
showing the presence of a strong shock front between 2 and 3 AU.
A factor of 100 increase in midplane gas density occurs when passing
through this shock front. This shock front is being driven by the
gravitational forces associated with the clumps and spiral arms that
have formed primarily between 5 AU and 10 AU in the disk. Note, however,
that even if these clumps do not succeed in forming giant planets,
their transient existence will still drive shock fronts at 2.5 AU,
permitting thermal processing of chondrule precursors from essentially
$t = 0$ for the solar nebula, the time at which CAIs formed some
4,567 Ma ago (Amelin et al. 2002). In addition, once Jupiter forms by either
core accretion or disk instability, it may continue to drive strong 
shock fronts in the inner disk similar to those seen in Fig.~1 for as
long as the inner disk gas remains. Thus, chondrule formation is
expected to proceed for several Ma, the estimated lifetime of
the inner solar nebula.

 The spiral pattern producing the inner shock fronts rotates at a
speed governed roughly by the Keplerian angular velocity of the clumps
orbiting beyond 5 AU. As a result, there can be a large difference
in orbital velocity between this spiral pattern and dust aggregates
moving with the gas on nearly Keplerian orbits at orbital radii inside (or
outside) the distances of the clumps (Wood 1996). For a spiral pattern
rotating with the angular velocity appropriate for a clump orbiting
at 5.2 AU, the orbital velocity difference is 12 km s$^{-1}$ between
this spiral pattern at 2.5 AU and solids orbiting with the gas at 2.5 AU.
However, it is only the component of this velocity difference
{\it perpendicular} to the shock front that is relevant for thermal
processing. Inner and outer shock fronts are usually so tightly wound 
that solids moving with
the gas through such shocks would strike them at a low angle and hence
with a velocity difference considerably less than 12 km s$^{-1}$. At the 
time shown in Fig.~1, however, the inner shock is oriented nearly obliquely
to the direction of motion of orbiting solids, at an angle of $\sim 60$
degrees, leading to a shock speed of $\sim$ 10 km s$^{-1}$, more than
sufficient for melting chondrule precursor dust aggregates. This transient
shock structure is the driving motivation for this paper. Most of the time
the angle around 2.5 AU is closer to $\sim 30$ degrees or smaller,
leading to a shock speed of $\sim$ 6 km s$^{-1}$ or less.
High-angle shocks have not been seen in the outer disk in this model.

 Multiple chondrule heating events may be required to produce
porphyritic texures in chondrules (Hewins \& Fox 2004), and transient though
persistent spiral waves lasting for several Ma are clearly
able to satisfy that constraint. However, the number of heating events
may also be limited by the need to preserve chondrule rims (Hewins 1996).
Global simulations of unstable disks integrated long enough for the
gravitational instabilities to settle into an asymptotic turbulent
behavior (Mej\'ia et al. 2005),
as well as the model presented in Fig.~1, show,
however, that shock fronts with the required conditions (shock speed,
pre- and post-shock densities) are by no means a permanent feature.
Rather, shocks which appear to have roughly the desired characteristics
are transients in a chaotic environment.
Moreover, it is possible that gravitational instabilities in the solar
nebula were intermittent due to the episodic build up of mass in
a ``dead zone'' where radial transport by magnetohydrodynamic turbulence was
inhibited (Gammie 1996; Armitage, Livio, \& Pringle 2001).
Episodes of rapid transport by gravitational instabilities may have
been associated with FU Orionis outbursts. Future work is needed to estimate
how often a protoplanetary disk is likely to produce a shock front
suitable for chondrule formation.

\section{Unified Scenario for Chondrules and CAIs}

 In Fig.~2 we present a unified scenario for the chronology of
the early Solar System and the thermal processing of solids. Models
of gravitationally unstable disks have shown (Boss 2004) that mixing and
transport of mm-sized and even cm-sized solids occurs on a short
time scale ($\sim 0.001$ Ma), faster than such solids could spiral
in to the protosun by gas drag and be lost. In addition, in a
marginally gravitationally unstable nebula of the type needed to
form Jupiter by either core accretion or disk instability, the
nebula will have transient spiral arms and rings that produce
local pressure maxima in the disk midplane. Because of these midplane
pressure maxima, solids will feel headwinds and tailwinds that
drive them toward the centers of the arms and rings (Haghighipour \& Boss
2003; Rice et al. 2004; Durisen et al. 2005).
This prevents the loss of solids by migration toward the protosun that would
otherwise result from gas drag due to headwinds in a nebula where the
gas pressure decreased monotonically with distance from the protosun.
Instead, the CAIs, chondrules, and smaller dust grains will remain tied to
the gas of the gravitationally evolving disk, and concentration of solids in
dense structures will accelerate formation of planetesimals,
planetary embryos, and asteroidal parent bodies through mutual collisions.
These bodies are large enough to be effectively decoupled
from the disk gas, and so remain behind after the inner disk gas is
depleted. In the magnetically ``dead zone'' model of solar nebula
evolution, chondrite parent-body forming events may occur episodically.
Within 3 Ma, the inner disk is largely cleared of
gas and primordial dust, and over the next $\sim 30$ Ma the
terrestrial planets then grow by collisional accumulation of the
planetesimals and planetary embryos previously formed there.

\section{Conclusions}

Spiral waves driven by gravitational instabilities appear to be
a natural, almost unavoidable means for thermally processing
chondrule precursors in the early solar nebula. Chondrules that
were formed in the asteroidal region of the solar nebula would have
been mixed together with CAIs that formed closer to the protosun and
with smaller, unheated grains (matrix) to form the chondritic meteorites.
The mixing, segregration by dense structures, and thermal processing
of chondritic material may thus be intimately connected with dynamic
processes in the gas disk that led to gas giant planet formation
in the outer nebula.

We thank A. C. Boley and the referee for useful comments. 
A.P.B.'s research is supported in part 
by the NASA Planetary Geology and Geophysics Program under grant NNG04GG15G
and the NASA Origins of Solar Systems Program under grant NAG5-11569.
Calculations were performed on the Carnegie Alpha Cluster, the purchase
of which was supported in part by NSF MRI grant AST-9976645.
R.H.D.'s research was supported in part by the NASA Origins of
Solar Systems Program grant NAG5-11964.

\vfill\eject

\begin{figure}
\vspace{-2.0in}
\plotone{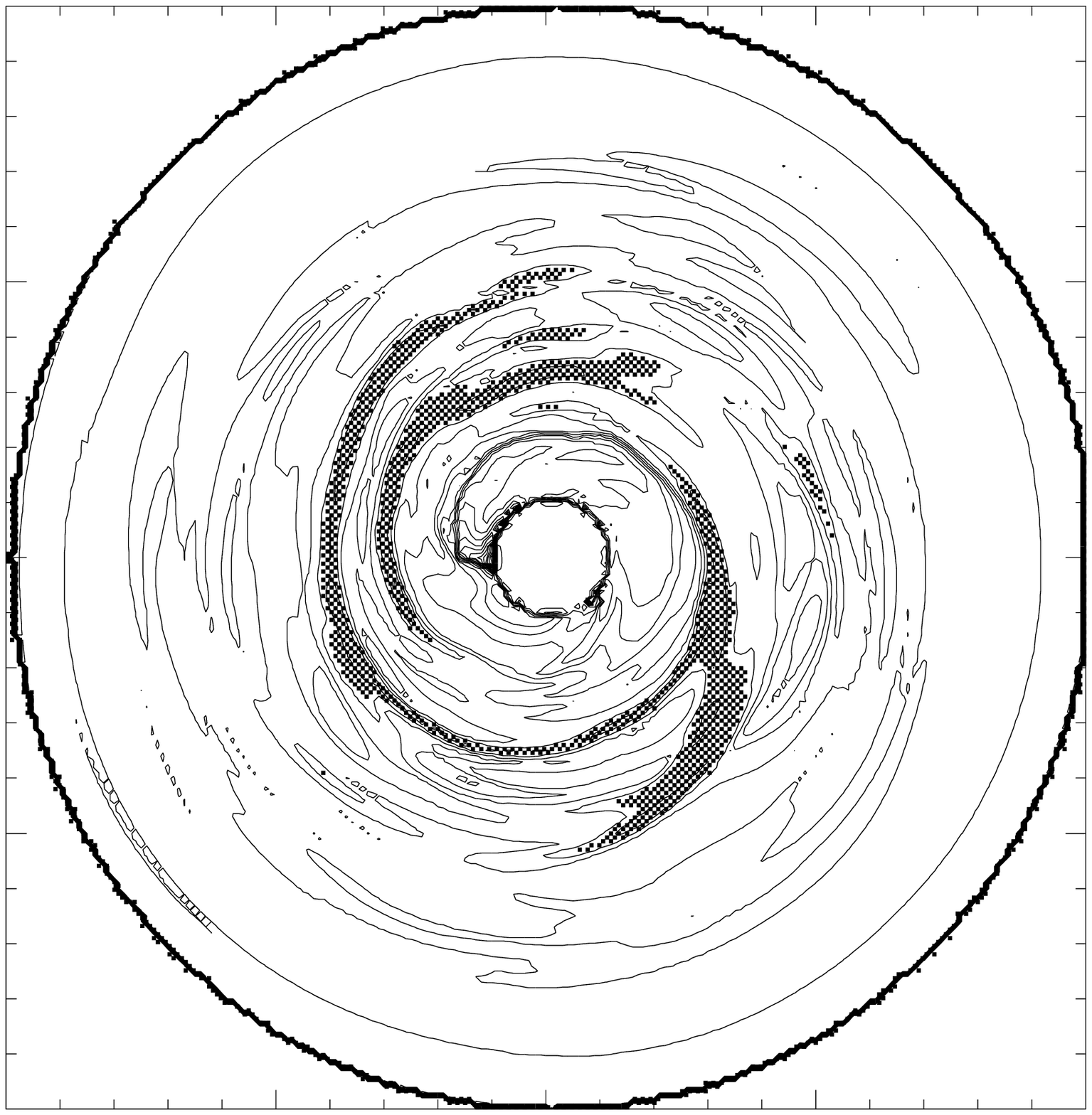}
\caption{Density contours in the midplane of a gravitationally unstable
disk after 252 years of evolution from an initially nearly axisymmetric
state, showing a strong transient shock front located at 9 o'clock,
just outside the inner boundary of radius 2 AU. Radius of the entire region
shown is 20 AU. Cross-hatched regions denote regions with densities
above $10^{-10}$ g cm$^{-3}$. A solar-mass protostar lies at the center
of the disk. Solids rotating in the counterclockwise
direction between 2 and 3 AU encounter the shock front at
a speed of $\sim 10$ km s$^{-1}$.}
\end{figure}

\begin{figure}
\caption{\small{A unified scenario for the evolution of solids in the inner 
solar nebula. (a) At $t = 0$ Ma, defined by the 4,567 Ma age of CAIs, CAIs are
present in the disk, formed close to the protosun and possibly lofted by the
protosun's bipolar outflow (Shu et al. 2001) to greater distances
(streamlines
above and below disk). The bulk of the disk is magnetically dead because
of the low ionization fraction, while the surface of the disk is ionized
and magnetically active (Gammie 1996).
(b) The disk is marginally gravitationally unstable, resulting in the
rapid inward and outward transport of CAIs and dust grains (Boss 2004).
(c) Spiral arms form Jupiter-mass clumps as disk mixing and transport
of solids continues.
(d) The spiral arms and clumps at 5 AU and beyond drive strong shock
fronts in the inner disk, capable of thermally processing precursor
dust aggregates into chondrules.
(e) Jupiter and Saturn form either rapidly (Boss 1997, 2001) or slowly
(Inaba et al. 2003)
but in either case continue to drive shock fronts intermittently
at asteroidal distances. Chondrules, CAIs, and matrix-sized dust grains
collide and form planetesimals and planetary embryos in the inner disk.
(f) Within 3 Ma, the inner solar nebula is accreted by the protosun,
leaving behind the rocky bodies that will collide over the next
$\sim 30$ Ma to form the terrestrial planets and the asteroid belt.}}
\end{figure}

\begin{figure}
\plotone{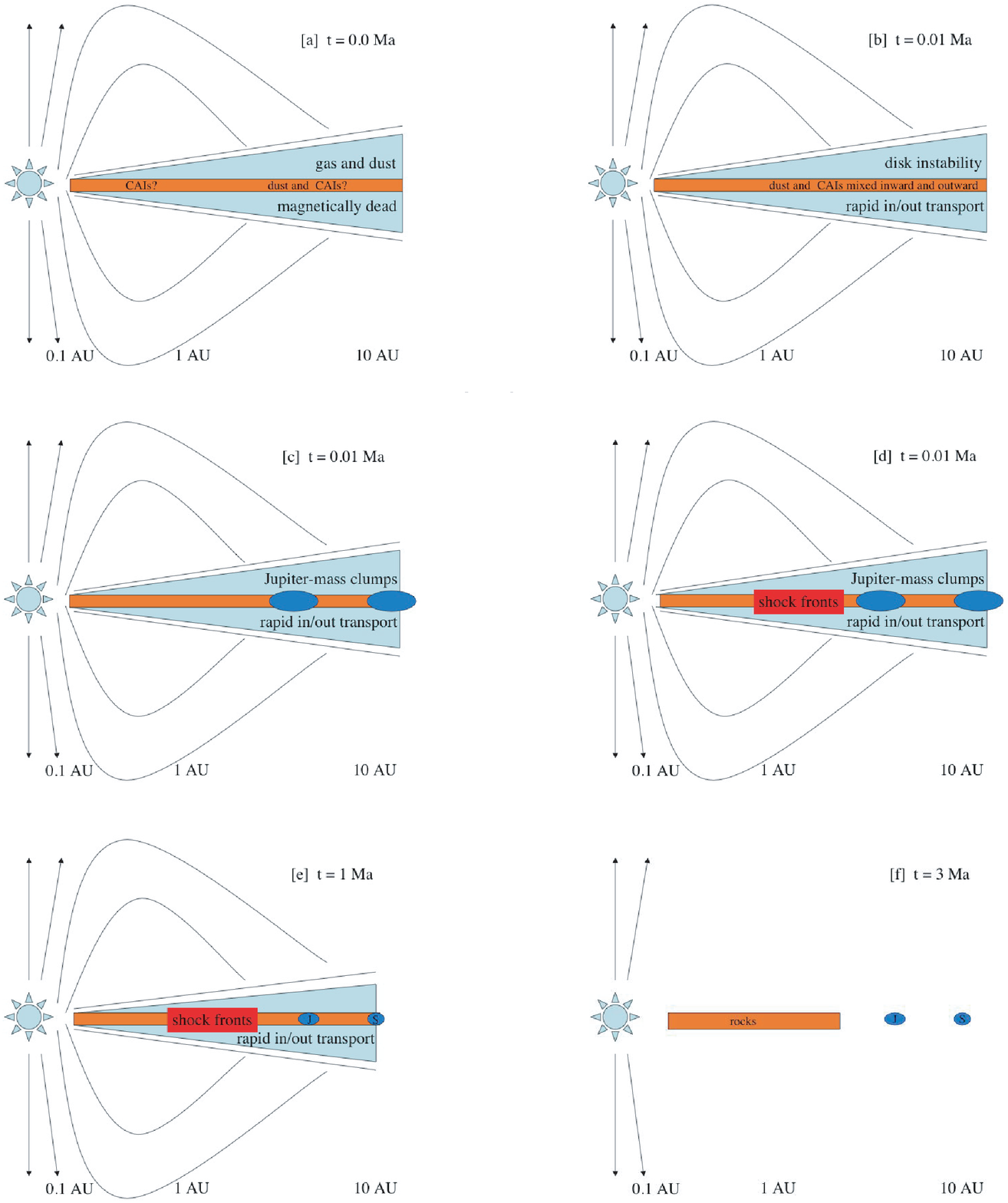}
\end{figure}

\end{document}